\documentstyle[prl,twocolumn,aps]{revtex}  
\def\circledR{^{\bigcirc\!\llap{\protect\tiny R}}\ } 

\catcode`\^^M=10

\begin{document}                  
\input epsf
\bibliographystyle{prsty}

\title{Prediction of Stable Walking for a Toy  That Cannot Stand} 

\author{Michael J. Coleman$^*$, 
        Mariano Garcia$^{*\dagger}$, 
        Katja Mombaur$^+$, and  
        Andy Ruina$^*$ }

\address{$^*$Department of Theoretical and Applied Mechanics,
             Cornell Univ., Ithaca, NY, 14853-7501 USA,mjc23@cornell.edu\\
         $^+$IWR - Univ. Heidelberg, Im Neuenheimer Feld 368, 
	69120 Heidelberg, D, katja.mombaur@iwr.uni-heidelberg.de\\
        $^\dagger$now at Ithaca Technical Center, Borg-Warner Automotive
                          770 Warren Rd. Ithaca NY 14850
\\
         {\em  For submission to PRE } 
        }

\date{Draft Aug 11, 2000,  revised \today}

\maketitle

\abstract{Previous  experiments [M. J. Coleman and A. Ruina,
Phys.~Rev.~Lett.~{\bf 80}, 3658 (1998)] showed that a
gravity-powered toy with no control and which has  no statically stable
near-standing configurations can walk stably.  We show here that a
simple rigid-body statically-unstable mathematical model based loosely
on the physical toy can predict stable limit-cycle walking motions.
These calculations add to the repertoire of  rigid-body
mechanism behaviors as well as further  implicating passive-dynamics
as a possible contributor to  stability of animal motions. }


\paragraph*{\bf Introduction.}

\par

For walking and other activities, people and animals move in complex,
yet stable ways.  One view is that such coordination is the action of
neuromuscular control constrained by, among other things, the laws of
classical mechanics. However, one might ask how much of animal
coordination might be understood in purely mechanical terms.  Likewise,
how much versatility of motion is possible with simple
mechanical devices?  This paper concerns one example that sheds a
little light on these two general questions.

\par 

McGeer's (e.g.~\cite{mcgeer3:1990}) success
with straight-legged, two-di\-men\-sio\-nal uncontrolled and gravity powered
walking mechanisms highlights the possibility of pure mechanics 
generating  coordination.  McGeer found steady walking solutions
(periodic gait or limit-cycle motions) that were exponentially stable
(asymptotically returned to the periodic motion after small disturbances from
that motion). In his two-dimensional theory, only fore-aft stability, and not
lateral balance, is in question. In his physical implementations side-to-side
balance is enforced  by duplicate side-by-side legs (4 legs total).  These
machines cannot stand fully upright, but can stand with splayed legs, possibly
contributing to their dynamic stability.

\par 
Extending McGeer's ideas, Coleman and Ruina~\cite{coleman2:1997} described an
easily reproducible~\cite{biscardi:1998} two legged gadget built from
Tinkertoys$\circledR$ that cannot stand at all, even with both feet on
the ground, splayed or not, yet seems (slightly) dynamically stable
while walking.  But where the stability of McGeer's machines was 
first predicted with rigid-body modeling, the stability of the
Tinkertoy$\circledR$ device was not.

As noted in~\cite{coleman2:1997}, this system is essentially dissipative (from
collisions and possibly from ground friction and internal dissipation) and
nonholonomic (the dimension of the accessible configuration space is larger
than the dimension of the  velocity space). Nonholonomic
systems can have asymptotic stability, even when conservative, and
nonholonomicity from intermittent foot contact might also contribute to
stability~\cite{ruina:1997,colemanholmes:1999}.

\par

Which properties are needed for asymptotic dynamic
stability of such a statically unstable system was left unanswered
by~\cite{coleman2:1997}.  Possible key factors 
include friction of the hinge, play in the hinge joint, elastic or
inelastic deformation of the structure, compliance at the foot contact,
and sliding and twisting friction at the foot contact. Could a rigid
body model without these effects explain the stable motion?

\begin{figure}
\centerline{\epsfxsize=3.5in\epsffile{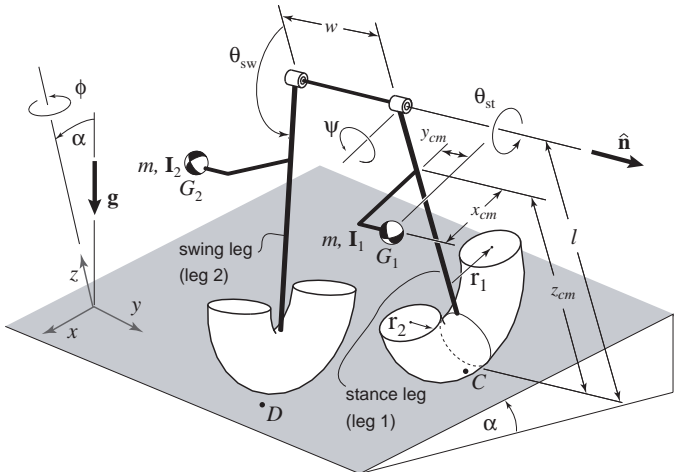}}
\caption{The 3D rigid body model.  The parameters and state variables
are described in~\protect\cite{coleman2:1997}.}
\label{3Dconfig}
\end{figure}

\paragraph*{\bf Previous research.} The simulation model
in~\cite{coleman2:1997} consists of two rigid bodies connected by a
frictionless hinge (Fig.~\ref{3Dconfig}).  The feet are toroidal with principal
radii $r_1$ and $r_2$.  The ground allows no
relative motion of contacting points, no torques at the foot contact, no
bounce (restitution $e=0$), and no tension  force at the foot contact
(non-sticky floor). The lengths, center of mass location, the
moment of inertia components, and the ground slope are adjustable. After
non-dimensionalizing with mass $m$, length
$l$ and time $\sqrt{l/g}$ there are 13 free parameters.

The working Tinkertoy$\circledR$ 
was based on mildly unstable simulations \cite{coleman2:1997} 
of a simplified model with point contact ($r_1=r_2=0$) and narrow hips ($w=0$).

\par 

Earlier, McGeer~\cite{mcgeer6:1992} studied the same model, allowing
$r_1>0$ and $w>0$ but assuming that the principal moments of inertia aligned
with the hip hinge and leg.  He found only unstable solutions where,
also, the swing foot passed below ground.

Kuo~\cite{kuo:1999} studied a similar model, but disallowed steer
($\phi$) and also only found unstable passive gaits.

Dankowicz~\cite{danko:1998} found stable solutions for a related 
 kneed computational model.  That model has wide feet so,
like  the 2D models, it can stand stably with splayed legs.
The semi-3D computational model of Wisse {\it et al.}~\cite{wisse:2000} can also
stand with splayed legs.

\paragraph*{\bf Methods.}
Our study was of the system in
(Fig.~\ref{3Dconfig})~\cite{coleman2:1997} described above, but with hip
spacing and disc feet ($r_1\ne 0, w\ne0$, $r_2=0$). 

\par
The overall approach is to characterize the solution of the rigid-body
dynamics equations as a function (map) that takes  the state of the system
just after one step as input and gives the state just after the next step as
output.  A fixed point of this map defines a limit cycle.  Stability is
evaluated by the eigenvalues of the matrix describing the linearization of
this map. If all eigenvalues are less than one in modulus the periodic motion
is exponentially stable. The map, its fixed points and the linearization are
all found from numerical solutions of the equations of classical rigid-body
dynamics.

\par
The numerical searches were aimed at generating stable motion and not
at accurately modeling either the existent physical toy or humans. We
used the toy's approximate parameters to seed the optimizations.
Special purpose optimal control software (see Appendix)
was used to reduce the most unstable eigenvalue while maintaining
periodicity of the gait, positive foot clearance and static
instability.  The resulting solution was checked and improved with 
an independent method and checked again with another independent simulation.

\par

\begin{figure}
\centerline{\epsfxsize=3.in \epsffile{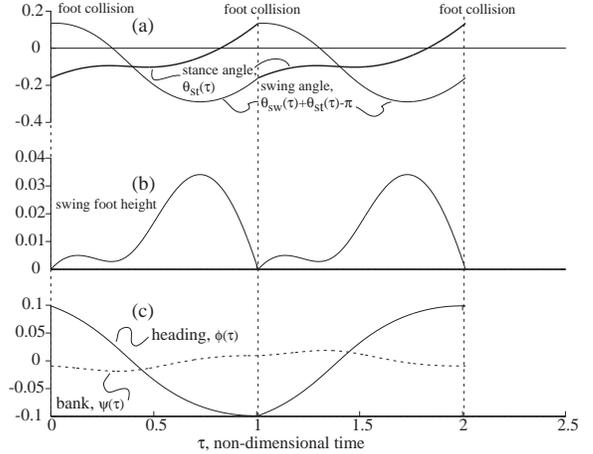}}
\caption{A gait cycle (two steps).  In the first half, the stance leg
is the left leg. (a) The swing leg angle
is here measured from the slope normal
($\theta_{sw}^{*}(\tau)+\theta_{st}^{*}(\tau)-\pi$); (b) positive
swing-foot clearance between collisions; (c) the motion has more steer (yaw)
than lean (bank).}
\label{leg.angles.paths}
\end{figure}

\paragraph*{\bf Results.} 
The model of Fig.~\ref{3Dconfig} has asymptotically stable limit cycle
motions (Fig.~\ref{leg.angles.paths}), with the foot of the swing leg
clearing the ground, with $I_{XX}=0.1982$, $I_{YY}=0.0186$,
$I_{ZZ}=0.1802$, $I_{XY}=0.0071$, $I_{XZ}=-0.0023$, $I_{YZ}=0.0573$,
$\alpha=0.0702$, $X_{cm}=0$, $Y_{cm}=0.6969$, and $Z_{cm}=0.3137$, $W
=.3624$, and $R_1= 0.1236$ and $R_2=0$.  Capital letters indicate
non-dimensional variables. Tensor components $I_{MN}$ and mass
positions are in local left-leg coordinates with origin at the
vertically-standing contact. Note the static instability
($Z_{cm}>R_1$).  The largest eigenvalue modulus of the single-step map
Jacobian is $0.8391560$, safely below 1.

\paragraph*{\bf Discussion}

We are claiming a qualitative theoretical mechanics result.  That
is,  a system described with the classic equations of rigid body
mechanics has an exponentially stable limit-cycle solution in the
neighborhood of a statically unstable configuration, but with no fast
spinning parts.  Although we have not attempted a mathematical proof,
we have attempted to do our numerics with enough checks and tests to state
the  result with confidence (see appendix below).

\par
This solution can be interpreted as  bipedal walking, although not
especially anthropomorphic.  The base solution is exactly repetitive,
step after step.  That the largest eigenvalue is less than 1 in
magnitude means that, if the mechanism were slightly disrupted from this
periodic motion it would asymptotically approach this motion again, over a
number of steps.

\par
Simple numerical probes show, as  exponential stability demands, a small but
non-infinitesimal  basin of attraction. We have not investigated the shape
or size of this domain in detail; we do not know exactly what set of
motions  eventually settle into the periodic motion  and thus cannot precisely
describe what disturbances  knock the walker down. However, the success of the
physical model~\cite{coleman2:1997} suggests that the stability is robust enough
to be physically relevant.

\par
We do not claim to have an accurate model of the toy in \cite{coleman2:1997}.
Rather we have a simple model that explains the toy's qualitative behavior.
Accurate quantitative prediction of the toy's motions may well depend on
physical effects that are not in our simple model (various frictions
and deformations).  Yet unknown is whether the parameters presented here could
be used as a basis for a better working  physical device.  More generally, we
also do not know if more human-like stable passive-dynamic
designs can be made which are also statically unstable.

\paragraph*{\bf Conclusions.}
The dynamic stability of a statically unstable walking mechanism can be
predicted with a model consisting of two rigid bodies connected by an
ideal hinge and making intermittent ideal no-slip, no-bounce point
contact with the ground. We have shown that there is no need to
appeal to hinge-friction, hinge-play, distributed or contact deformation 
(elastic or inelastic), or contact frictional slip in order to
qualitatively predict the interesting behavior demonstrated by the physical model
in~\cite{coleman2:1997}. 

\par
 The results further highlight the versatility of simple passive
strategies for stabilization of coordination. The calculation also slightly
expands the range of known rigid-body phenomenology.

\par For videos and reprints about the Tinkertoy$\circledR$ and
related machines, see www.tam.cornell.edu/$^\sim$ruina/hplab/pdw.html.

\paragraph*{\bf Acknowledgments.}  This work was
supported by a biomechanics grant from the National Science Foundation and a
travel grant from Heidelberg University. We thank Hans-Georg Bock for encouraging
this collaboration. 

\appendix 
\paragraph*{\bf Appendix: Numerical Analysis} 
We carried out the numerical analysis three different ways. 

\par
The first stable solution (with $|\sigma_{max}|=0.897$)  with foot
clearance was found using the approach  developed by Mombaur {\it et al.}
\cite{mombaur:2000} on the basis of the optimal control code Muscod by Bock {\it et
al.}\cite{bock:1984} and Leineweber \cite{leineweber:1996}. In the language
of the discipline, Muscod has been written for general multi-phase optimal control
problems and is based on a multiple shooting state discretization. Multiple
shooting splits up the original boundary value problem into a number of initial
value problems enforcing  continuity conditions at the transitions from one
interval to the next. At all multiple shooting points Muscod allows the user to
impose  a number of equality and inequality  constraints on the parameters
and the dynamic variables being varied  in the optimization. For the tinkertoy
model described here,  this permitted us to ensure periodicity, foot clearance
during the step, and to keep all state variables and
parameters within reasonable ranges.

Sensitivities of the integration end values on each interval, both to variations
in initial values and to variations in model parameters,
are efficiently computed  by means  of internal numerical differentiation
(IND).  The basic principle of IND is  to use identical, but adaptive and
error-controlled    discretization schemes for integration and derivative
generation.
 
For use with actuated and passive  gait problems, with implicit state-dependent
phase switching points  and discontinuities in the state variables, the original
Muscod has been combined with an object oriented modeling library that deals 
with these situations in a uniform and complete way. Also added to Muscod were 
stability analysis modules which compute the linearized
Poincar{\'e} map of a periodic solution  and their eigenvalues, assembling
information from all multiple shooting intervals and taking care of the above
mentioned implicit switching point dependencies.

Stable solutions for the tinkertoy model were found by varying model  parameters
and bounds based on coarse  grid sensitivity information gathered during the
previous optimal control problem solutions.

\par

Second, we reproduced and improved  the solution above by 4th order Runge-Kutta
integration of the ODE's in Matlab${\circledR}$, finding the collision time
accurately using Henon's method (changing the independent variable near the
collision time). The impact jump transition is a matrix multiplication.  The fixed
points of the resulting return map were found by numerical root finding.  The
fixed point map Jacobian was found by a central difference perturbation of the
initial state.  The eigenvalue was  reduced from 0.897 to 0.839 using a simulated
annealing optimization of the maximum eigenvalue modulus. For this Matlab solution
we did extensive convergence tests on both the integration step size and the
central difference step size. These tests indicate a combined
roundoff and truncation error  of about $\pm 10^{-7}$ in the
largest eigenvalue modulus for the parameters given. This maximum
eigenvalue estimate differs  from that generated by  Muscod with
these parameters by $2\times 10^{-3}$.

\par

Finally, the equations of motion were derived independently and simulated
again independently in Matlab giving agreement to the
Matlab solution above to  $10^{-6}$ for the largest eigenvalue
modulus.

\par
For reference, the state of the system just after
collision of the left foot is, for the parameters given, 
${\bf q}^{*}
  = [\phi, \psi, \theta_{st}, \theta_{sw},
     \dot\phi, \dot\psi,\dot\theta_{st},\dot\theta_{sw} ]$   
 $=[0.09868, $
 $-0.00925, $ 
 $-0.16016, $
 $ 3.43583, $
 $-0.13221, $
 $-0.01991, $ 
 $ 0.47124, $
 $-0.39256]$
with  step period $\tau^*=1.00711$ where 
$\dot{(\,\,\, )}=d(\,\,\,)/d\tau$ with
$\tau$ the dimensionless time.



\end{document}